\documentclass{revtex4}
\usepackage{amssymb}
\usepackage{amsmath}

\input{tcilatex}

\begin{document}

\title[ ]{Black hole solutions in f(R) gravity coupled with non-linear
Yang-Mills field }
\author{S. Habib Mazharimousavi}
\email{habib.mazhari@emu.edu.tr}
\author{M. Halilsoy}
\email{mustafa.halilsoy@emu.edu.tr}
\affiliation{Department of Physics, Eastern Mediterranean University, G. Magusa, north
Cyprus, Mersin 10, Turkey. }

\begin{abstract}
It is shown that in the static, spherically symmetric spacetime the problem
of metric $f(R)$ gravity coupled with non-linear Yang-Mills (YM) field
constructed from the Wu-Yang ansatz as source, can be solved in all
dimensions. By non-linearity it is meant that the YM Lagrangian depends
arbitrarily on its invariant. A particular form is considered to be in the
power-law form with limit of the standard YM theory. The formalism admits
black hole solutions with single or double horizons in which $f(R)$ can be
obtained, in general numerically. In $6-$dimensional case we obtain an exact
solution given by $f(R)=\sqrt{R}$ gravity that couples with the YM field in
a consistent manner.

\emph{We dedicate this work to the memory of Yavuz Nutku (1943-2010), the
renowned mathematical physicist from Turkey.}
\end{abstract}

\maketitle

\section{Introduction}

With the hopes to explain a number of cosmological problems covering dark
energy, accelerated expansion, quantum gravity and many related matters,
extensions / modifications of general relativity theory gained momentum anew
during the recent decade. Each extension adds new degrees of freedom and
accommodates new parameters apt for the sake of better physics. Lovelock
gravity \cite{1}, for instance, constitutes one such extension which abides
by the ghost free combinations of higher order invariants resulting in
second order equations alone. Next higher order to the Einstein-Hilbert
extension in this hierarchy came to be known as the Gauss-Bonnet extension 
\cite{2} which makes use of the quadratic invariants. Apart from this
hierarchy, arbitrary dependence on the Ricci scalar $R$ which has been
popular in recent times is known as the $f(R)$ gravity (see e.g. \cite{3}
and references therein and for a review paper see \cite{4}). Compared to
other theories which employ tensorial invariants this sounds simpler and the
fact that the ghosts are eliminated makes it attractive \cite{5}. The
simplest form, namely $f(R)=R$ is the well-known Einstein-Hilbert Lagrangian
which constitutes the simplest theory of gravity. Given the simplest theory
in hand, why to investigate complex versions of it? The idea is to add new
degrees of freedom through non-linearities, create curvature sources that
may be counterbalanced by the energy-momentum of some physical sources. Once 
$f(R)=R$, passes all the classical experimental tests any function $f(R)$
may be interpreted as a self-similar version of $f(R)=R$, creating no
serious difficulties at the classical level. Introducing an effective
Newtonian constant as a matter of fact plays a crucial role in this matter.
At the quantum level, however, problems such as unitarity, renormalizability
of the linearized theory are of vital importance to be tackled with.

As an example we refer to the particular form $f(R)=R^{N}$ ($N=$ rational
number) \cite{6}. This admits, among others an exact solution which
simulates the geometry of a charged object i.e. the Reissner-Nordstr\"{o}m
geometry \cite{6}. That is, $f(R)=R^{N}$ behaves geometrically as if we have 
$f(R)=R+$ (electrostatic field). Remarkably, the power $N$ plays the role of
"charge" so that the geometry $f(R)=R^{N}$ becomes locally isometric to the
geometry of Reissner-Nordstr\"{o}m. In a similar manner various combinations
of polynomial forms plus $ln\left( 1+R\right) $, $sinR$, and other functions
of $R$ can be considered as potential candidates for $f(R)$. Some of these
have already appeared in the literature \cite{7}. Beside the case of $%
f(R)=R+ $ (electrostatic field), and more aptly, cases such as $f(R)=R+$
(non-minimal scalar field) cases also have been investigated \cite{7}.

Recently, the non-minimal Yang-Mills fields coupled with $f(R)$ gravity has
also been studied \cite{8}. In this paper we show that Yang-Mills (YM) field
can be accommodated within the context of metric $f(R)$ gravity as well. To
the best of our knowledge such a study, especially in higher dimensions
which constitutes our main motivation, is absent in the literature. While
numerical solutions to the problem of black holes in $f(R)-$YM theory \cite%
{9} started to appear in the literature our interest is in finding exact
solutions. It should be added also that the class of black holes in $f(R)$
gravity can be distinct from the well-known classes such as Myers-Perry (in 
\cite{2}). Some classes of black holes in this paper also obey this rule
since they are not asymptotically flat in the usual sense. We show that in
most of our solutions asymptotically (i.e. $r\rightarrow \infty $) an
effective cosmological constant can be identified which depends on the
dimension of spacetime, YM charge $Q$ and the integration parameters.
Herein, we do not propose an $f(R)$ Lagrangian a priori, instead we
determine the $f(R)$ function in accordance with the YM sources. By using
the Wu-Yang ansatz for the YM field \cite{10,11} it is shown that a general
class of solutions can be obtained in the $f(R)$ gravity where the geometric
source matches with the energy-momentum of the YM field. Let us note that
the Wu-Yang ansatz works miraculously in all higher dimensions which renders
possible to solve $f(R)$ gravity not only in $d=4$, but in all $d>4$ as
well.\ Further, the zero trace condition for the energy-momentum is imposed
to obtain conformal invariant solutions which constitutes a particular
class. Although our starting point is the non-linear YM field in which the
Lagrangian is an arbitrary function of the YM invariant our main concern is
the special limit, namely, the linear (or standard) YM theory. Power-law
type non-linearity is a well-known class which we consider as an example and
as we have shown elsewhere \cite{12} the choice of the power plays a crucial
role in the satisfaction of the energy conditions, i.e. Weak, Strong or
Dominant. Essentially, conformal invariant property is one of the reasons
that we consider Power-YM (PYM) field in higher dimensions. The
implementation of zero trace condition for the energy-momentum tensor
becomes relatively simpler in this PYM class. As in the example of
Born-Infeld electrodynamics case which plays crucial role in resolution of
point like singularities, in analogy, similar expectations can be associated
with the non-linear version of the YM theory. Such non-linearities resemble
the self-interacting scalar fields which serve to define different vacua in
quantum field theory. More to that, in general relativity the non-linear
terms effect black hole formation significantly, it is therefore tempting to
take such combinations seriously. It is our belief that with the non-linear
YM field we can establish effective cosmological parameters to contribute,
in accordance with the energy conditions cited above, to the distinction
between the phantom and quintessence data of our universe. This is a
separate problem of utmost importance that should be considered separately.
We show that, the choice $f(R)=\sqrt{R}$ in $6-$dimensions yields an exact
solution for $f(R)$ gravity coupled with YM fields which is
non-asymptotically flat / non-de Sitter in the sense that it contains
deficit angles at $r\rightarrow \infty .$ Other classes of solutions that
are asymptotically de Sitter, unfortunately can't be expressed in a closed
form as $f(R).$\ 

Organization of the paper is as follows. In Section II we introduce our
theory of non-linear YM field coupled to $f(R)$ gravity and give exact
solutions. Section III specifies the non-linearity of YM field to PYM case
in all dimensions. The First Law of thermodynamics in our formalism is
discussed briefly in Sec. IV. We complete the paper with Conclusion which
appears in Sec. V.

\section{Non-linear YM field in $f(R)$ gravity}

We start with an action given by%
\begin{equation}
S=\int d^{d}x\sqrt{-g}\left[ \frac{f\left( R\right) }{2\kappa }+L\left(
F\right) \right]
\end{equation}%
in which $f\left( R\right) $ is a real function of Ricci scalar $R$, $%
L\left( F\right) $ is the non-linear YM Lagrangian with $F=\frac{1}{4}%
tr\left( F_{\mu \nu }^{\left( a\right) }F^{\left( a\right) \mu \nu }\right)
. $ The particular choice $L\left( F\right) =-\frac{1}{4\pi }F$ will reduce
to the case of standard YM theory. Here 
\begin{equation}
\mathbf{F}^{\left( a\right) }=\frac{1}{2}F_{\mu \nu }^{\left( a\right)
}dx^{\mu }\wedge dx^{\nu }
\end{equation}%
is the YM field $2-$form with the internal index $(a)$ running over the
degrees of freedom of the YM non-abelian gauge field. Our unit convention is
chosen such that $c=G=1$ so that $\kappa =8\pi .$

Variation of the action with respect to the metric gives the field equations
as%
\begin{equation}
f_{R}R_{\mu }^{\nu }+\left( \square f_{R}-\frac{1}{2}f\right) \delta _{\mu
}^{\nu }-\nabla ^{\nu }\nabla _{\mu }f_{R}=\kappa T_{\mu }^{\nu }
\end{equation}%
where $f_{R}=\frac{df\left( R\right) }{dR}$ and $\square f_{R}=\nabla _{\mu
}\nabla ^{\mu }f_{R}=\frac{1}{\sqrt{-g}}\partial _{\mu }\left( \sqrt{-g}%
\partial ^{\mu }\right) f_{R}.$ Further, $\nabla ^{\nu }\nabla _{\mu
}f_{R}=g^{\alpha \nu }\left( f_{R}\right) _{,\mu ;\alpha }=g^{\alpha \nu }%
\left[ \left( f_{R}\right) _{,\mu ,\alpha }-\Gamma _{\mu \alpha }^{m}\left(
f_{R}\right) _{,m}\right] .$\ The trace of the field equation implies%
\begin{equation}
f_{R}R+\left( d-1\right) \square f_{R}-\frac{d}{2}f=\kappa T
\end{equation}%
in which $T=T_{\mu }^{\mu }.$

The energy momentum tensor is chosen to be%
\begin{equation}
T_{\mu }^{\nu }=L\left( F\right) \delta _{\mu }^{\nu }-tr\left( F_{\mu
\alpha }^{\left( a\right) }F^{\left( a\right) \nu \alpha }\right)
L_{F}\left( F\right)
\end{equation}%
in which $L_{F}\left( F\right) =\frac{dL\left( F\right) }{dF}.$

The YM ansatz, following the higher dimensional extension of Wu-Yang ansatz,
is given by%
\begin{align}
\mathbf{A}^{(a)}& =\frac{Q}{r^{2}}C_{\left( i\right) \left( j\right)
}^{\left( a\right) }\ x^{i}dx^{j},\text{ \ \ }Q=\text{YM magnetic charge, \ }%
r^{2}=\overset{d-1}{\underset{i=1}{\sum }}x_{i}^{2}, \\
2& \leq j+1\leq i\leq d-1,\text{ \ and \ }1\leq a\leq \left( d-2\right)
\left( d-1\right) /2,  \notag \\
x_{1}& =r\cos \theta _{d-3}\sin \theta _{d-4}...\sin \theta _{1},\text{ }%
x_{2}=r\sin \theta _{d-3}\sin \theta _{d-4}...\sin \theta _{1},  \notag \\
\text{ }x_{3}& =r\cos \theta _{d-4}\sin \theta _{d-5}...\sin \theta _{1},%
\text{ }x_{4}=r\sin \theta _{d-4}\sin \theta _{d-5}...\sin \theta _{1}, 
\notag \\
& ...  \notag \\
x_{d-2}& =r\cos \theta _{1},  \notag
\end{align}%
in which $C_{\left( b\right) \left( c\right) }^{\left( a\right) }$ is the
non-zero structure constants \cite{10}. The spherically symmetric metric is
written as%
\begin{equation}
ds^{2}=-A\left( r\right) dt^{2}+\frac{dr^{2}}{A\left( r\right) }%
+r^{2}d\Omega _{d-2}^{2},
\end{equation}%
where $A\left( r\right) $ is the only unknown function of $r$ and 
\begin{equation}
d\Omega _{d-2}^{2}=d\theta _{1}^{2}+\underset{i=2}{\overset{d-2}{\tsum }}%
\underset{j=1}{\overset{i-1}{\tprod }}\sin ^{2}\theta _{j}\;d\theta _{i}^{2},
\end{equation}%
with%
\begin{equation*}
0\leq \theta _{d-2}\leq 2\pi ,0\leq \theta _{i}\leq \pi ,\text{ \ \ }1\leq
i\leq d-3.
\end{equation*}%
The YM equations take the form 
\begin{equation}
\mathbf{d}\left[ ^{\star }\mathbf{F}^{\left( a\right) }L_{F}\left( F\right) %
\right] +\frac{1}{\sigma }C_{\left( b\right) \left( c\right) }^{\left(
a\right) }L_{F}\left( F\right) \mathbf{A}^{\left( b\right) }\wedge ^{\star }%
\mathbf{F}^{\left( c\right) }=0,
\end{equation}%
where $^{\star }$ means duality. For our future use we add also that 
\begin{equation}
F=\frac{1}{4}tr\left( F_{\mu \nu }^{\left( a\right) }F^{\left( a\right) \mu
\nu }\right) =\frac{\left( d-2\right) \left( d-3\right) Q^{2}}{4r^{4}}
\end{equation}%
and%
\begin{equation}
tr\left( F_{t\alpha }^{\left( a\right) }F^{\left( a\right) t\alpha }\right)
=tr\left( F_{r\alpha }^{\left( a\right) }F^{\left( a\right) r\alpha }\right)
=0,
\end{equation}%
while%
\begin{equation}
tr\left( F_{\theta _{i}\alpha }^{\left( a\right) }F^{\left( a\right) \theta
_{i}\alpha }\right) =\frac{\left( d-3\right) Q^{2}}{r^{4}}.
\end{equation}%
From (5) the non-zero energy momentum tensor components are%
\begin{eqnarray}
T_{t}^{t} &=&L=T_{r}^{r}, \\
T_{\theta _{i}}^{\theta _{i}} &=&L-\frac{\left( d-3\right) Q^{2}}{r^{4}}%
L_{F}.
\end{eqnarray}%
The trace of Eq. (5) implies%
\begin{equation}
T=d.L-4FL_{F}
\end{equation}%
which yields from Eq. (3)%
\begin{equation}
f=\frac{2}{d}\left[ f_{R}R+\left( d-1\right) \square f_{R}-\kappa \left(
d.L-4FL_{F}\right) \right] .
\end{equation}%
To write the exact form of the field equations we need the general form of
Ricci scalar and Ricci tensor which are given by%
\begin{eqnarray}
R &=&-\frac{r^{2}A^{\prime \prime }+2\left( d-2\right) rA^{\prime }+\left(
d-2\right) \left( d-3\right) \left( A-1\right) }{r^{2}}, \\
R_{t}^{t} &=&R_{r}^{r}=-\frac{1}{2}\frac{rA^{\prime \prime }+\left(
d-2\right) A^{\prime }}{r}, \\
R_{\theta _{i}}^{\theta _{i}} &=&-\frac{rA^{\prime }+\left( d-3\right)
\left( A-1\right) }{r^{2}}.
\end{eqnarray}%
in which a prime denotes derivative with respect to $r$. Overall, the field
equations read now%
\begin{eqnarray}
f_{R}\left( -\frac{1}{2}\frac{rA^{\prime \prime }+\left( d-2\right)
A^{\prime }}{r}\right) +\left( \square f_{R}-\frac{1}{2}f\right) -\nabla
^{t}\nabla _{t}f_{R} &=&\kappa L, \\
f_{R}\left( -\frac{1}{2}\frac{rA^{\prime \prime }+\left( d-2\right)
A^{\prime }}{r}\right) +\left( \square f_{R}-\frac{1}{2}f\right) -\nabla
^{r}\nabla _{r}f_{R} &=&\kappa L, \\
f_{R}\left( -\frac{rA^{\prime }+\left( d-3\right) \left( A-1\right) }{r^{2}}%
\right) +\left( \square f_{R}-\frac{1}{2}f\right) -\nabla ^{\theta
_{i}}\nabla _{\theta _{i}}f_{R} &=&\kappa \left( L-\frac{4}{\left(
d-2\right) }FL_{F}\right) .
\end{eqnarray}%
Herein 
\begin{equation}
\square f_{R}=\frac{1}{\sqrt{-g}}\partial _{r}\left( \sqrt{-g}\partial
^{r}\right) f_{R}=A^{\prime }f_{R}^{\prime }+Af_{R}^{\prime \prime }+\frac{%
\left( d-2\right) }{r}Af_{R}^{\prime },
\end{equation}%
\begin{equation}
\nabla ^{t}\nabla _{t}f_{R}=g^{tt}f_{R,t;t}=g^{tt}\left( f_{R,t,t}-\Gamma
_{tt}^{m}f_{R,m}\right) =\frac{1}{2}A^{\prime }f_{R}^{\prime },
\end{equation}%
\begin{equation}
\nabla ^{r}\nabla _{r}f_{R}=g^{rr}f_{R,r;r}=g^{rr}\left( f_{R,r,r}-\Gamma
_{rr}^{m}f_{R,m}\right) =Af_{R}^{\prime \prime }+\frac{1}{2}A^{\prime
}f_{R}^{\prime },
\end{equation}%
and%
\begin{equation}
\nabla ^{\theta _{i}}\nabla _{\theta _{i}}f_{R}=g^{\theta _{i}\theta
_{i}}f_{R,\theta _{i};\theta _{i}}=g^{\theta _{i}\theta _{i}}\left(
f_{R,\theta _{i},\theta _{i}}-\Gamma _{\theta _{i}\theta
_{i}}^{m}f_{R,m}\right) =\frac{A}{r}f_{R}^{\prime }.
\end{equation}%
The $tt$ and $rr$ components of the field equations imply%
\begin{equation}
\nabla ^{r}\nabla _{r}f_{R}=\nabla ^{t}\nabla _{t}f_{R}
\end{equation}%
or equivalently%
\begin{equation}
f_{R}^{\prime \prime }=0.
\end{equation}%
This leads to the solution%
\begin{equation}
f_{R}=\xi +\eta r
\end{equation}%
where $\xi $ and $\eta $ are two integration constants and to avoid any
non-physical case we assume that $\eta ,\xi >0$. The other field equations
become%
\begin{eqnarray}
\left( \xi +\eta r\right) \left( -\frac{1}{2}\frac{rA^{\prime \prime
}+\left( d-2\right) A^{\prime }}{r}\right) +\frac{1}{2}\eta A^{\prime }+%
\frac{\left( d-2\right) }{r}A\eta -\frac{1}{2}f &=&\kappa L, \\
\left( \xi +\eta r\right) \left( -\frac{rA^{\prime }+\left( d-3\right)
\left( A-1\right) }{r^{2}}\right) +A^{\prime }\eta +\frac{\left( d-3\right) 
}{r}A\eta -\frac{1}{2}f &=&\kappa \left( L-\frac{4}{\left( d-2\right) }%
FL_{F}\right) .
\end{eqnarray}%
In a similar manner the $\theta _{i}\theta _{i}$ and $tt$ components yield%
\begin{equation}
\left( \xi +\eta r\right) \left( \frac{2\left( d-3\right) \left( A-1\right)
-r^{2}A^{\prime \prime }-\left( d-4\right) rA^{\prime }}{2r^{2}}\right)
+\left( \frac{A}{r}-\frac{A^{\prime }}{2}\right) \eta =\kappa \frac{4}{%
\left( d-2\right) }FL_{F}.
\end{equation}%
Equation (16) can equivalently be expressed by%
\begin{equation}
f=\frac{2}{d}\left[ f_{R}R+\left( d-1\right) \left( A^{\prime }\eta +\frac{%
\left( d-2\right) }{r}A\eta \right) -\kappa \left( d.L-4FL_{F}\right) \right]
.
\end{equation}

Here we comment that in the limit of linear Einstein-YM (EYM) theory one may
set $L=-\frac{1}{4\pi }F,L_{F}=-\frac{1}{4\pi },\eta =0,$ and $\xi =1$ to
get $f_{R}=1$ or equivalently $f=R$ and consequently%
\begin{equation}
2\left( d-3\right) \left( A-1\right) -r^{2}A^{\prime \prime }-\left(
d-4\right) rA^{\prime }=-\frac{4\left( d-3\right) Q^{2}}{r^{2}}.
\end{equation}%
This admits a solution as%
\begin{equation}
A\left( r\right) =\left\{ 
\begin{array}{cc}
1-\frac{m}{r^{d-3}}-\frac{d-3}{d-5}\frac{Q^{2}}{r^{2}}, & d>5 \\ 
1-\frac{m}{r^{2}}-\frac{2Q^{2}\ln r}{r^{2}}, & d=5%
\end{array}%
\right.
\end{equation}%
which was reported before \cite{11}. Here we note that $m$ is an integration
constant related to mass of the black hole.

\section{PYM field coupled to $f\left( R\right) $ gravity}

\subsection{General integral for the PYM field in $f(R)$ gravity}

Our first approach to the solution of the field equations, concerns the PYM
theory which is a particular non-linearity given by the Lagrangian $L=-\frac{%
1}{4\pi }F^{s},$ in which $s$ is a real parameter \cite{13}. The EYM limit
is obtained by setting $s=1.$

The metric function, then, reads as ($\xi =0$)%
\begin{equation}
A\left( r\right) =\left\{ 
\begin{array}{cc}
\frac{d-3}{d-2}+\Lambda r^{2}-\frac{m}{r^{d-2}}-\frac{\left( d-1\right)
\left( d-2\right) ^{\frac{d-1}{2}}\left( d-3\right) ^{\frac{d-1}{4}}}{2^{%
\frac{d-5}{2}}\eta d}\frac{Q^{\frac{d-1}{2}}\ln r}{r^{d-2}}, & s=\frac{d-1}{4%
} \\ 
\frac{d-3}{d-2}+\Lambda r^{2}-\frac{m}{r^{d-2}}-\frac{4^{2-s}s\left(
d-2\right) ^{s-1}\left( d-3\right) ^{s}Q^{2s}}{\left( 4s+1\right) \left(
d-4s-1\right) \eta r^{4s-1}}, & s\neq \frac{d-1}{4}%
\end{array}%
\right.
\end{equation}%
in which $\Lambda $ and $m$ arise naturally as integration constants.
Obviously $\Lambda $ is identified as the cosmological constant while $m$ is
related to the mass. The fact that our metric is asymptotically de Sitter
seems to be manifest only with deficit angles at $r\rightarrow \infty .$ We
add that in order to have an exact solution we had to set $\xi =0,$ which
means that in this case we were unable to obtain the $f(R)=R$ gravity from
the general solution.

Using the metric solution we also find $f\left( R\left( r\right) \right) $
and $R\left( r\right) $ as%
\begin{equation}
f\left( R\left( r\right) \right) =\left\{ 
\begin{array}{cc}
2\eta \frac{d-3}{r}-\frac{\left( d-3\right) ^{\frac{d-1}{4}}\left(
d-2\right) ^{\frac{d-5}{4}}}{2^{\frac{d-5}{2}}}\frac{Q^{\frac{d-1}{2}}}{%
r^{d-1}}, & s=\frac{d-1}{4} \\ 
2\eta \frac{d-3}{r}-\frac{\left( d-3\right) ^{s}\left( d-2\right)
^{s-1}\left( 4s-d+2\right) }{4^{s-1}}\frac{Q^{2s}}{r^{4s}}, & s\neq \frac{d-1%
}{4}%
\end{array}%
\right.
\end{equation}%
and 
\begin{equation}
R\left( r\right) =\left\{ 
\begin{array}{cc}
\frac{d-3}{r^{2}}-\Lambda d\left( d-1\right) -\frac{\left( d-1\right) \left(
d-3\right) ^{\frac{d-1}{4}}\left( d-2\right) ^{\frac{d-5}{4}}}{2^{\frac{d-5}{%
2}}\eta d}\frac{Q^{\frac{d-1}{2}}}{r^{d}}, & s=\frac{d-1}{4} \\ 
\frac{d-3}{r^{2}}-\Lambda d\left( d-1\right) -\frac{\left( 4s-d+2\right)
s\left( d-3\right) ^{s}\left( d-2\right) ^{s-1}}{4^{s-2}\left( 4s+1\right)
\eta }\frac{Q^{2s}}{r^{4s+1}}, & s\neq \frac{d-1}{4}%
\end{array}%
\right. .
\end{equation}%
We recall from equation (29) that%
\begin{equation}
f_{R}=\frac{df}{dR}=\eta r
\end{equation}%
or equivalently%
\begin{equation}
\frac{df/dr}{dR/dr}=\eta r.
\end{equation}%
As it is seen from the expressions of $R\left( r\right) $ and $f\left(
r\right) $ it is not possible to eliminate $r$ to have the exact form of $%
f\left( R\right) ,$ instead we have a parametric form for $f\left( R\right) $%
.

Among all possible cases, we are interested in the condition $4s-d+2=0.$
Since this particular choice brings significant simplifications in (37) and
(38). Table 1 shows for which values of $s$ and $d$ this is satisfied.%
\begin{equation}
\begin{tabular}{|l|l|l|l|l|l|l|l|}
\hline
$d=$ & $5$ & $6$ & $7$ & $8$ & $9$ & $10$ & $d$ \\ \hline
$s=$ & $\frac{3}{4}$ & $1$ & $\frac{5}{4}$ & $\frac{3}{2}$ & $\frac{7}{4}$ & 
$2$ & $\frac{d-2}{4}$ \\ \hline
\end{tabular}%
.  \tag{Table 1}
\end{equation}%
It is not difficult to observe that with these specific choices, plus $%
\Lambda =0,$ we obtain%
\begin{equation}
f\left( R\right) =\mu _{\circ }\sqrt{R}
\end{equation}%
in which the constant $\mu _{\circ }$ is defined by%
\begin{equation}
\mu _{\circ }=2\eta \sqrt{d-3}.
\end{equation}%
Accordingly the metric function $A\left( r\right) $ takes the form 
\begin{equation}
A\left( r\right) =\frac{d-3}{d-2}-\frac{m}{r^{d-2}}-\frac{\left( d-2\right)
^{\frac{d-2}{4}}\left( d-3\right) ^{\frac{d-2}{4}}Q^{\frac{d-2}{2}}}{4^{%
\frac{d-6}{4}}\left( d-1\right) \eta r^{d-3}}
\end{equation}%
with the scalar curvature%
\begin{equation}
R\left( r\right) =\frac{d-3}{r^{2}}.
\end{equation}%
Since the constant term $\frac{d-3}{d-2}\neq 1,$ in (43) our solution for $%
r\rightarrow \infty $ is given by 
\begin{equation}
ds^{2}\overset{r\rightarrow \infty }{=}-d\bar{t}^{2}+d\bar{r}^{2}+\left(
\alpha \bar{r}\right) ^{2}d\Omega _{d-2}^{2}
\end{equation}%
where $\bar{t}=\sqrt{\alpha }t,$ $r=\alpha \bar{r},$ ($\alpha =\frac{d-3}{d-2%
}$). This may be interpreted as a deficit angle at $r\rightarrow \infty .$
It can also be seen easily from Table 1 that for the linear YM theory ($s=1$%
) in $d=6$, with $\eta =\frac{\sqrt{3}}{6},$ $f(R)=\sqrt{R}$ yields an exact
solution.

\subsection{Thermodynamics of the black hole solution}

The black hole solution given by (36) admits horizon(s) provided%
\begin{equation}
A\left( r_{h}\right) =0,
\end{equation}%
which implies%
\begin{equation}
m=\left\{ 
\begin{array}{cc}
\frac{d-3}{d-2}r_{h}^{d-2}+\Lambda r_{h}^{d}-\frac{\left( d-1\right) \left(
d-2\right) ^{\frac{d-1}{2}}Q^{\frac{d-1}{2}}\left( d-3\right) ^{\frac{d-1}{4}%
}}{2^{\frac{d-5}{2}}\eta d}\ln r_{h}, & s=\frac{d-1}{4} \\ 
\frac{d-3}{d-2}r_{h}^{d-2}+\Lambda r_{h}^{d}-\frac{4^{2-s}s\left( d-2\right)
^{s-1}\left( d-3\right) ^{s}Q^{2s}}{\left( 4s+1\right) \left( d-4s-1\right)
\eta r_{h}^{4s-d+1}}, & s\neq \frac{d-1}{4}%
\end{array}%
\right. .
\end{equation}%
The standard definition of Hawking temperature 
\begin{equation}
T_{H}=\frac{1}{4\pi }A^{\prime }\left( r_{h}\right)
\end{equation}%
yields%
\begin{equation}
T_{H}=\left\{ 
\begin{array}{cc}
\frac{\frac{\eta }{4}d\left( d-3\right) ^{1/4}\left( \left( r_{h}^{2}\Lambda
+1\right) d-3\right) -\sqrt{2}\left( d-2\right) ^{\frac{d-5}{4}}\left(
d-1\right) r_{h}^{1-d}Q^{\frac{d-1}{2}}\left( \frac{d-3}{4}\right) ^{\frac{d%
}{4}}r_{h}}{\left( d-3\right) ^{\frac{1}{4}}\pi r_{h}\eta d}, & s=\frac{d-1}{%
4} \\ 
\frac{\left[ \eta \left( 4s+1\right) \left( d-2\right) \left( r^{2}\Lambda
d+d-3\right) -4\left( \frac{d-2}{4}\right) ^{s}\left( d-3\right)
^{s}Q^{2s}sr_{h}^{1-4s}\right] }{4\pi \eta \left( d-2\right) \left(
4s+1\right) r_{h}}, & s\neq \frac{d-1}{4}%
\end{array}%
\right. .
\end{equation}%
It is known that the area formula $S=\frac{\mathcal{A}_{H}}{4G},$ in $f(R)$
gravity becomes \cite{14,15,16}%
\begin{equation}
S=\frac{\mathcal{A}_{h}}{4G}\left. f_{R}\right\vert _{r=r_{h}}
\end{equation}%
in which 
\begin{equation}
f_{R}=\eta r_{h}
\end{equation}%
and%
\begin{equation}
\mathcal{A}_{h}=\frac{d-1}{\Gamma \left( \frac{d+1}{2}\right) }\pi ^{\frac{%
d-1}{2}}r_{h}^{d-2}
\end{equation}%
where $r_{h}$ is the radius of the event horizon or cosmological horizon of
the black hole. Using $S$ with the definition of the heat capacity in
constant charge we get%
\begin{eqnarray}
C_{Q} &=&T_{H}\left( \frac{\partial S}{\partial T_{H}}\right) _{Q}= \\
&&\frac{\pi ^{\frac{d-1}{2}}\eta \left( d-1\right) ^{2}r^{d-1}}{4\Gamma
\left( \frac{d+1}{2}\right) }\frac{\eta \left( d-2\right) \left( s+\frac{1}{4%
}\right) \left( \Lambda r^{2}d+d-3\right) -4s\left( \frac{d-2}{4}\right)
^{s}\left( d-3\right) ^{s}Q^{2s}r^{-4s+1}}{\eta \left( d-2\right) \left( s+%
\frac{1}{4}\right) \left( \Lambda r^{2}d-d+3\right) +16s^{2}\left( \frac{d-2%
}{4}\right) ^{s}\left( d-3\right) ^{s}Q^{2s}r^{-4s+1}}.  \notag
\end{eqnarray}%
A through analysis of the zeros / infinities of this function reveal about
local thermodynamic stability / phase transitions, which will be ignored
here.

\subsection{A general approach with $s=1$}

\subsubsection{$d\geq 6$}

In this section, for the linear YM theory ($s=1$), we let $\xi $ to get
non-zero value and attempt to find the general solution. As one may notice
the case of $d=5$ is distinct so that we shall find a separate solution for
it but for $d\geq 6$ the most general solution reads%
\begin{eqnarray}
A\left( r\right) &=&1-\frac{m}{r^{d-3}}-\frac{\left( d-3\right) Q^{2}}{%
\left( d-5\right) \xi r^{2}}+\eta \Delta _{1}+\eta ^{2}\Delta _{2}+\eta
^{3}\Delta _{3}+\eta ^{4}\Delta _{4}+  \notag \\
&&\left( -1\right) ^{d}r^{2}\gamma ^{d-1}\left( d-1\right) m\ln \left\vert 
\frac{\xi +\eta r}{r}\right\vert +P_{d-7}\left( \gamma \right) ,
\end{eqnarray}%
in which $\gamma =\frac{\eta }{\xi }$ with the following abbreviations%
\begin{eqnarray}
\Delta _{1} &=&\frac{4\left( d-3\right) }{3\left( d-5\right) }\frac{Q^{2}}{%
\xi ^{2}r}+\frac{d-1}{d-2}\frac{m}{\xi r^{d-4}}-\frac{2}{d-2}\frac{r}{\xi },
\notag \\
\Delta _{2} &=&\frac{2}{\left( d-2\right) \left( d-5\right) }\frac{r^{2}}{%
\xi ^{2}}\ln \left\vert \frac{\xi +\eta r}{r}\right\vert -\frac{d-1}{d-3}%
\frac{m}{\xi ^{2}r^{d-5}}-\frac{2\left( d-3\right) }{\left( d-5\right) }%
\frac{Q^{2}}{\xi ^{3}},  \notag \\
\Delta _{3} &=&\frac{4\left( d-3\right) }{d-5}\frac{Q^{2}}{\xi ^{4}}r+\frac{%
d-1}{d-4}\frac{m}{\xi ^{3}r^{d-6}},  \notag \\
\Delta _{4} &=&-\frac{4\left( d-3\right) }{d-5}\frac{Q^{2}}{\xi ^{5}}%
r^{2}\ln \left\vert \frac{\xi +\eta r}{r}\right\vert -\frac{d-1}{d-5}\frac{1%
}{r^{d-7}}, \\
P_{d-7}\left( \gamma \right) &=&\left( -1\right) ^{d-1}\left( d-1\right)
m\gamma ^{5}\left[ \overset{d-6\text{ term(s)}}{\overbrace{\gamma ^{d-7}r-%
\frac{1}{2}\gamma ^{d-8}+\frac{1}{3}\gamma ^{d-9}r^{-1}-...+\frac{1}{d-6}%
\gamma ^{-\left( d-8\right) }}}\right] \text{, \ \ \ \ \ \ \ \ \ }d\geq 7. 
\notag
\end{eqnarray}%
Using these results, we plot Fig. 1 which displays (for $d=6$), $A(r)$ and $%
f(R)$ for different values of $\eta $.

It can easily be seen that in the limit $\eta \rightarrow 0$ and $\xi =1,$
the metric function reduces to%
\begin{equation}
A\left( r\right) =1-\frac{m}{r^{d-3}}-\frac{\left( d-3\right) Q^{2}}{\left(
d-5\right) r^{2}},
\end{equation}%
which is nothing but the well-known black hole solution in $f(R)=R,$ EYM
theory \cite{11}. On the other hand for $\eta \neq 0\neq \xi ,$ it is
observed that asymptotical flatness does not hold.

To complete our solution we find the asymptotic behavior of the metric
function $A(r)$. As one observes from (54), at $r\rightarrow \infty $ $%
A\left( r\right) $ becomes 
\begin{equation}
A\left( r\right) \simeq 1+\frac{\Lambda _{eff}}{3}r^{2},
\end{equation}%
in which 
\begin{equation}
\Lambda _{eff}=3\left[ \frac{2}{\left( d-2\right) \left( d-5\right) }-\frac{%
4\left( d-3\right) }{d-5}Q^{2}\eta ^{4}+\left( -1\right) ^{d}\gamma
^{d-1}\left( d-1\right) \xi ^{2}m\right] \gamma ^{2}\ln \left\vert \eta
\right\vert .
\end{equation}%
As one may see in Fig. 1, we add here that $\eta $ plays a crucial role in
making the metric function asymptotically de Sitter, anti - de Sitter and
asymptotically flat ($\left\vert \eta \right\vert <1$, $\left\vert \eta
\right\vert >1$ and $\left\vert \eta \right\vert =1$ respectively).

\paragraph{Thermodynamics of the BH solution in $6-$dimensions}

In this part we would like to study the thermodynamics of the solution (54)
and compare the result with the case of linear gravity $f(R)=R.$ As one can
see from the form of the solution (54), we are not able to study
analytically in any arbitrary dimensions $d,$ and therefore we only consider 
$d=6$. The metric solution in $d=6$ dimensions is given by%
\begin{eqnarray}
A\left( r\right) &=&1-\frac{m}{r^{3}}-\frac{3Q^{2}}{\xi r^{2}}+\eta \left( 
\frac{4Q^{2}}{\xi ^{2}r}+\frac{5}{3}\frac{m}{\xi r^{2}}-\frac{1}{2}\frac{r}{%
\xi }\right) + \\
&&\eta ^{2}\left( \frac{1}{2}\frac{r^{2}}{\xi ^{2}}\ln \left\vert \frac{\xi
+\eta r}{r}\right\vert -\frac{5}{2}\frac{m}{\xi ^{2}r}-6\frac{Q^{2}}{\xi ^{3}%
}\right) +  \notag \\
&&\eta ^{3}\left( 12\frac{Q^{2}}{\xi ^{4}}r+\frac{5}{2}\frac{m}{\xi ^{3}}%
\right) +\eta ^{4}\left( -12\frac{Q^{2}}{\xi ^{5}}r^{2}\ln \left\vert \frac{%
\xi +\eta r}{r}\right\vert -5r\right) ,  \notag
\end{eqnarray}%
and therefore the Hawking temperature is given by%
\begin{equation}
T_{H}=\left( \frac{3}{4\pi r_{h}}-\frac{3}{4}\frac{Q^{2}}{\pi r_{h}^{3}}%
\right) +\left( \frac{17Q^{2}}{16\pi r_{h}^{2}}-\frac{3}{16}\frac{1}{\pi }%
\right) \eta -\left( \frac{123Q^{2}}{64\pi r_{h}}+\frac{120r_{h}\ln
r_{h}+139r_{h}}{192\pi }\right) \eta ^{2}+\mathcal{O}\left( \eta ^{3}\right)
\end{equation}%
and the specific heat capacity reads%
\begin{equation}
C_{Q}=\left( \frac{8\pi ^{2}r_{h}^{4}\left( Q^{2}-r_{h}^{2}\right) }{\left(
3r_{h}^{2}-9Q^{2}\right) }\right) +\left( -\frac{8}{9}\frac{\pi
^{2}r_{h}^{5}\left( 3r_{h}^{4}-17Q^{2}r_{h}^{2}+7Q^{4}\right) }{\left(
r_{h}^{2}-3Q^{2}\right) ^{2}}\right) \eta +\mathcal{O}\left( \eta
^{2}\right) .
\end{equation}%
First we comment herein that, to get the above result we considered $\xi =1.$
Second we add that, by $\eta =0$ we get the case of EYM black hole in $R-$%
gravity. In the case of pure $R-$gravity we put $\eta =0$ and $Q=0$ which
leads to%
\begin{equation}
T_{H}=\frac{3}{4\pi r_{h}},\text{ and }C_{Q}=-\frac{8\pi ^{2}}{3}r_{h}^{4}
\end{equation}%
which are the Hawking temperature and Heat capacity of the $6-$dimensional
Schwarzschild black hole. Divergence in the Heat capacity for particular YM
charge and therefore a thermodynamic instability is evident from this
expression.

\subsubsection{$d=5$}

As we stated before, dimension $d=5$ behaves different from the other
dimensions. The metric function is given in this case by%
\begin{equation}
A\left( r\right) =1-\frac{m}{r^{2}}-\frac{2Q^{2}}{\xi r^{2}}\ln r+\eta
\Delta _{1}+\eta ^{2}\Delta _{2}+\eta ^{3}\Delta _{3}+\eta ^{4}\Delta _{4}
\end{equation}%
in which%
\begin{eqnarray}
\Delta _{1} &=&\frac{1}{9\xi ^{2}r}\left[ 24Q^{2}\ln r+12m+2Q^{2}-6\xi r^{2}%
\right] ,  \notag \\
\Delta _{2} &=&\frac{1}{3\xi ^{3}}\left[ 2\xi r^{2}\ln \frac{\xi +\eta r}{r}%
-12Q^{2}\ln r-3Q^{2}-6m\xi \right] ,  \notag \\
\Delta _{3} &=&\frac{2r}{\xi ^{4}}\left[ 2m\xi -3Q^{2}+4Q^{2}\ln r\right] ,
\\
\Delta _{4} &=&-\frac{2r^{2}}{\xi ^{5}}\left[ 4Q^{2}\ln r\ln \frac{\xi +\eta
r}{\xi \sqrt{r}}+4Q^{2}\text{dilog}\left( \frac{\xi +\eta r}{\xi }\right)
+\left( 2m\xi -Q^{2}\right) \ln \left( \frac{\xi +\eta r}{r}\right) \right] .
\notag
\end{eqnarray}%
Herein $m$ is an integration constant and 
\begin{equation}
\text{dilog}\left( x\right) =\int\limits_{1}^{x}\frac{\ln t}{1-t}dt
\end{equation}%
is the dilogarithm function. Here also the EYM limit with $\eta \rightarrow
0 $ and $\xi =1$ is obvious.

Similar to the higher than $6-$dimensional case we give here also the
asymptotic behavior of the metric solution (65) as $r\rightarrow \infty ,$%
\begin{equation}
A\left( r\right) \simeq 1+\frac{\Lambda _{eff}}{3}r^{2},
\end{equation}%
in which 
\begin{equation}
\Lambda _{eff}=-2\frac{\eta ^{4}}{\xi ^{5}}\left\{ 6\left[ 3m\xi ^{5}\ln
\eta -Q^{2}\ln \left( \frac{\eta }{\xi }\right) \ln \left( \xi \eta \right) %
\right] -\frac{\xi ^{3}}{\eta ^{2}}\ln \eta +2\pi ^{2}Q^{2}\right\}
\end{equation}%
is the effective cosmological constant.

\subsection{Black holes with a conformally invariant YM source}

One of the interesting choice for $s$ in Einstein-Power-Maxwell theory -
which has been considered first by Hassaine and Martinez \cite{13} - is
given by $s=\frac{d}{4}$ (for all $d\geq 4$) which is conformally invariant.
This choice yields a zero trace for the energy momentum tensor in any
dimensions, i.e., $T=T_{\mu }^{\mu }=0$. In EPYM case also $s=\frac{d}{4}$
leads to a traceless energy momentum tensor and a metric solution for the
field equations with arbitrary values of $\xi $ and $\eta $ is given by%
\begin{equation}
A\left( r\right) =1-\frac{m}{r^{d-3}}+4\left( d-2\right) ^{\frac{d}{4}%
}\left( \frac{d-3}{4}\right) ^{\frac{d}{4}}\frac{Q^{\frac{d}{2}}}{\xi r^{d-2}%
}+r^{2}\left[ \frac{2\gamma ^{2}}{d-2}+\left( -1\right) ^{d}\left(
d-1\right) m\gamma ^{d-1}\right] \ln \left\vert \frac{\xi \left( 1+\gamma
r\right) }{r}\right\vert -\frac{2\gamma }{d-2}r+q\left( \gamma \right)
\end{equation}%
in which 
\begin{equation}
q\left( \gamma \right) =\left( -1\right) ^{d}\left( d-1\right) m\gamma
^{d-2}\sum\limits_{k=1}^{d-2}\frac{\left( -1\right) ^{k}r^{2-k}}{\gamma
^{k-1}k},
\end{equation}%
and $\gamma =\frac{\eta }{\xi }.$ Fig. 2 ($d=5,s=\frac{5}{4}$) and Fig. 3 ($%
d=6,s=\frac{3}{2}$) depict $A(r),R(r)$ and $f(R)$ which relate the
conformally invariant ($s=\frac{d}{4}$) cases for different $\eta $ values.
For $r\rightarrow \infty $, it can be seen easily from Eq. (68) that we have
an effective cosmological constant term, given by 
\begin{equation}
\Lambda _{eff}=3\left[ \frac{2\gamma ^{2}}{d-2}+\left( -1\right) ^{d}\left(
d-1\right) m\gamma ^{d-1}\right] \ln \left\vert \eta \right\vert
\end{equation}%
We note that as a limit, once $\eta \rightarrow 0$ (or equivalently $\gamma
\rightarrow 0$) and $\xi \rightarrow 1$ the solution reduces to%
\begin{equation}
A\left( r\right) =1-\frac{m}{r^{d-3}}+4\left( d-2\right) ^{\frac{d}{4}%
}\left( \frac{d-3}{4}\right) ^{\frac{d}{4}}\frac{Q^{\frac{d}{2}}}{r^{d-2}}
\end{equation}%
which is the metric function in Einstein-PYM theory in $f(R)=R$ gravity.
Determination of horizons and thermodynamical properties in this limit is
much more feasible in comparison with the intricate expression (68). To
complete this section we give the Hawking temperature and specific heat
capacity for $d=5$ which read%
\begin{equation}
T_{H}=\frac{2r_{h}^{3}-\sqrt[4]{24Q^{10}}}{4\pi r_{h}^{4}}+\frac{2\sqrt[4]{%
24Q^{10}}-r_{h}^{3}}{6\pi r_{h}^{3}}\eta -\frac{17r_{h}^{3}+10\sqrt[4]{%
24Q^{10}}+12r_{h}^{3}\ln r_{h}}{18\pi r_{h}^{2}}\eta ^{2}+\mathcal{O}\left(
\eta ^{3}\right)
\end{equation}%
and%
\begin{equation}
C_{Q}=-\frac{3}{4}\frac{\pi ^{2}r_{h}^{3}\left( \sqrt[4]{24Q^{10}}%
-2r_{h}^{3}\right) }{2\sqrt[4]{24Q^{10}}-r_{h}^{3}}+\frac{3\pi
^{2}r_{h}^{4}\left( 4\sqrt[4]{24Q^{10}}r_{h}^{3}-r_{h}^{6}-2\sqrt{6}%
Q^{5}\right) }{\left( 2\sqrt[4]{24Q^{10}}-r_{h}^{3}\right) }\eta +O\left(
\eta ^{2}\right) .
\end{equation}

\subsubsection{Constant $d-$dimensional curvature $R=R_{0}$}

G. Cognola, et al in Ref. \cite{15} have considered the constant
four-dimensional curvature $R=R_{0}$ in pure $f\left( R\right) $ gravity,
which implies a de-Sitter universe. Here we wish to follow the same
procedure in higher dimensions in $f\left( R\right) $ gravity coupled with
the non-minimall PYM field. As stated before, in order to have a traceless
energy momentum in $d-$dimensions we need to consider the case of
conformally invariant YM source which is given by $s=\frac{d}{4}$ in the PYM
source. In $4-$dimensions $s=1$ is satisfied automatically for the zero
trace condition.

We start with the trace of the Eq. (4) which leads to%
\begin{equation}
f^{\prime }\left( R_{0}\right) =\frac{d}{2R_{0}}f\left( R_{0}\right)
\end{equation}%
and therefore the field equations (3) become%
\begin{equation}
G_{\mu }^{\nu }+\Lambda _{eff}\delta _{\mu }^{\nu }=\kappa \text{\ }\tilde{T}%
_{\mu }^{\nu }
\end{equation}%
with the effective cosmological constant and energy momentum tensor as 
\begin{equation}
\Lambda _{eff}=\frac{\left( d-2\right) R_{0}}{2d},\text{ \ \ }\tilde{T}_{\mu
}^{\nu }=\frac{2R_{0}}{f\left( R_{0}\right) d}T_{\mu }^{\nu }\text{\ .}
\end{equation}%
Now, we follow \cite{15,16} to give the form of the entropy akin to the
possible BH solution. As we indicated in Eq. (50) the entropy of the
modified gravity with constant curvature is given by%
\begin{equation}
S=\frac{\mathcal{A}_{h}}{4G}f_{R_{0}}
\end{equation}%
which after considering (78) it becomes%
\begin{equation}
S=\frac{\mathcal{A}_{h}d}{8GR_{0}}f\left( R_{0}\right) .
\end{equation}%
Since our main concern in this paper is not the particular class of $%
R=R_{0}= $constant curvature space time we shall not extend our discussion
here any further.

\section{First Law of Thermodynamics}

In this section we follow Ref. \cite{17} to find a higher dimensional form
of the Misner-Sharp energy \cite{18} inside the horizon of the static
spherically symmetric black hole in $f(R)$ gravity. The corresponding metric
is given by (7) and the horizon is found from $A\left( r_{h}\right) =0.$ The
field equations (3) may be written as%
\begin{equation}
G_{\mu }^{\nu }=\kappa \left[ \frac{1}{f_{R}}T_{\mu }^{\nu }+\frac{1}{\kappa 
}\hat{T}_{\mu }^{\nu }\right]
\end{equation}%
where $G_{\mu }^{\nu }$ is the Einstein tensor and $\hat{T}_{\mu }^{\nu }$
is a stress-energy tensor for the effective curvature which reads%
\begin{equation}
\hat{T}_{\mu }^{\nu }=\frac{1}{f_{R}}\left[ \nabla ^{\nu }\nabla _{\mu
}f_{R}-\left( \square f_{R}-\frac{1}{2}f+\frac{1}{2}R\right) \delta _{\mu
}^{\nu }\right] .
\end{equation}%
At the horizon $tt$ and $rr$ parts of (79) imply%
\begin{equation}
\frac{d-2}{2r_{h}}A^{\prime }f_{R}-\frac{\left( d-2\right) \left( d-3\right) 
}{2r_{h}^{2}}f_{R}=\kappa \left( T_{0}^{0}+\frac{1}{2\kappa }\left[ \left(
f-Rf_{R}\right) -A^{\prime }f_{R}^{\prime }\right] \right)
\end{equation}%
which upon multiplying by an infinitesimal displacement $dr_{h}$ on both
sides can be reexpressed in the form 
\begin{equation}
\frac{A^{\prime }}{4\pi }d\left( \frac{2\pi \mathcal{A}_{h}}{\kappa }%
f_{R}\right) -\frac{1}{2\kappa }\left[ \frac{\left( d-2\right) \left(
d-3\right) }{r_{h}^{2}}f_{R}+\left( f-Rf_{R}\right) \right] \mathcal{A}%
_{h}dr_{h}=\mathcal{A}_{h}T_{0}^{0}dr_{h}.
\end{equation}%
We add here that all functions are calculated at the horizon, for instance $%
A^{\prime }=\left. \frac{dA\left( r\right) }{dr}\right\vert _{r=r_{h}}$. The
latter equation suggests that%
\begin{equation}
dE=\frac{1}{2\kappa }\left[ \frac{\left( d-2\right) \left( d-3\right) }{%
r_{h}^{2}}f_{R}+\left( f-Rf_{R}\right) \right] \mathcal{A}_{h}dr_{h}
\end{equation}%
in which $E$ is the Misner-Sharp energy in our case. Therefore (82) becomes%
\begin{equation}
TdS-dE=PdV
\end{equation}%
where we set Hawking temperature $T=\frac{A^{\prime }}{4\pi }$, entropy of
the black hole $S=\frac{2\pi \mathcal{A}_{h}}{\kappa }f_{R},$ radial
pressure of matter fields at the horizon $P=T_{r}^{r}=T_{0}^{0}$ and finally
the change of volume of the black hole at the horizon is given by $dV=%
\mathcal{A}_{h}dr_{h}.$ The exact form of the Misner-Sharp energy stored
inside the horizon may be found as%
\begin{equation}
E=\frac{1}{2\kappa }\int \left[ \frac{\left( d-2\right) \left( d-3\right) }{%
r_{h}^{2}}f_{R}+\left( f-Rf_{R}\right) \right] \mathcal{A}_{h}dr_{h}
\end{equation}%
in which the integration constant is set to zero (to read more see Ref.s 
\cite{17,19}).

As an example we study the case of PYM field in $f(R)$ gravity in Sec.
III-A. Also to have an exact form for $f(R)$ we employ the metric (43) which
corresponds to $f(R)=\mu _{\circ }\sqrt{R}.$ Eq. (82) yields,%
\begin{equation}
\frac{A^{\prime }}{4\pi }d\left( \frac{2\pi \mathcal{A}_{h}}{\kappa }\eta
r_{h}\right) -\frac{1}{2\kappa }\left[ \frac{\left( d-2\right) \left(
d-3\right) }{r_{h}}\eta +\frac{\eta \left( d-3\right) }{r_{h}}\right] 
\mathcal{A}_{h}dr_{h}=\mathcal{A}_{h}\left( \frac{-1}{4\pi }\left[ \frac{%
\left( d-2\right) \left( d-3\right) Q^{2}}{4r_{h}^{4}}\right] ^{\frac{d-2}{4}%
}\right) dr_{h},
\end{equation}%
in which $R=\frac{d-3}{r_{h}^{2}}$ has been used. Now, this equation leads to%
\begin{equation}
A^{\prime }=\frac{\left( d-3\right) }{r_{h}}-\frac{1}{\eta \left( d-1\right) 
}\left( 4\left[ \frac{\left( d-2\right) \left( d-3\right) Q^{2}}{4r_{h}^{4}}%
\right] ^{\frac{d-2}{4}}\right) .
\end{equation}%
By taking derivative of (43) and substituting for $m$ in terms of $r_{h}$
the foregoing equation easily follows.

Finally, one can see that the total energy is expressed by 
\begin{equation}
E=\frac{\eta \left( d-3\right) \left( d-1\right) }{2\kappa \left( d-2\right) 
}\mathcal{A}_{h}.
\end{equation}

\section{Conclusion}

An arbitrary dependence on the Ricci scalar in the form of $f(R)$ as
Lagrangian yields naturally an arbitrary geometrical curvature. The
challenge is to find a suitable energy-momentum that will match this
curvature by solving the highly non-linear set of equations. For a number of
reasons it has been suggested that $f(R)$ gravity may solve the
long-standing problems such as, accelerated expansion and dark energy
problems of cosmology. Richer theoretical structure naturally provides more
parameters to fit recent observational data. We have shown that in analogy
with the electromagnetic (both linear and non-linear) field, the Yang-Mills
field also can be employed and solved within the context of $f(R)$ gravity.
So far, $f(R)$ as a modified theory of gravity has been considered mainly in 
$d=4$, whereas we have been able in the presence of YM fields to extend it
to $d>4$. In addition to the parameters of the theory the dimension of space
time also contribute asymptotically to the effective cosmological constant
created in $f(R)$ gravity. Admittedly, out of the general numerical solution
technically it is not possible to invert scalar curvature $R(r)$ as $r=r(R)$
and obtain $f(R)$ in a closed form. This happens only in very special cases.
In particular dimensions and non-linearities we obtained black holes with
single / multi horizons. From the obtained solutions for PYM field coupled $%
f(R)$ gravity we can discriminate three broad classes as follows:

i) the asymptotically flat class in which $\eta =0,$ $\xi =1.$ This class
was already known \cite{11}.

ii) the asymptotically de-Sitter / anti-de Sitter class corresponding to $%
\eta \neq 0,$ $\xi =1$ ($s=1$).

iii) the non-asymptotically flat / non-asymptotically de Sitter class for $%
\eta \neq 0\neq \xi ,$ $s=\frac{d}{4}.$

Our solutions admit black hole solutions with single / multi horizons. In
the proper limits we recover all the well-known metrics to date. The case
(ii) at large distance limit exhibits deficit angle as shown in Eq. (45).

Conformally invariant class with zero trace of the energy-momentum tensor,
is obtained with the PYM Lagrangian $L\left( F\right) =-\frac{1}{4\pi }%
(F)^{5/4}$ in $d=5$. In general, the power of $F$ becomes meaningful within
the context of energy conditions and causality. By introducing effective
pressure $P_{eff}$ and energy density $\rho _{eff}$ through $P_{eff}=\omega
\rho _{eff}$ and using the PYM fields in energy conditions $\omega $ factor
(i.e. whether $\omega <-1,$ or $\omega >-1$) can be determined as a
cosmological factor \cite{4}. This will be our next project in this line of
study. It may happen that, certain set of powers eliminate non-physical
fields such as phantoms and alikes. As far as exact solutions are concerned
a remarkable solution is obtained in the case of standard YM Lagrangian $%
L\left( F\right) =-\frac{1}{4\pi }F$ with $d=6$ in $f(R)=\sqrt{R}$ gravity
which automatically restricts the curvature to $R>0.$

\textbf{ACKNOWLEDGEMENT:} We wish to thank T. Dereli, M. G\"{u}rses and B.
Tekin for much valuable discussions.

\textbf{Captions:}

\textbf{Fig. 1:} The plot of $6-$dimensional $f(R)$ (Fig. 1a) and $A(r)$
(Fig. 1b). We choose $\Lambda =0$, and four different values of $\eta $ ($%
\eta _{A},\eta _{B},\eta _{c}$ and $\eta _{D}$) are depicted as plots A, B,
C and D. From Fig. 1b it can be seen that in A, B and C we have single,
while in D double horizons.

\textbf{Fig. 2:} The $5-$dimensional plots of $A(r)$, $f(r)$ and $R(r)$ from
Eq. (68), for a variety of parameters given in Fig.s (2a-2d). Since $s=\frac{%
5}{4}$ in this particular case, the source is the PYM field with Lagrangian $%
L\sim F^{5/4}$. These are all black hole solutions with inner and outer
horizons. A general analytic expression for $f(R)$ seems out of our reach.

\textbf{Fig. 3: }The plot of the metric function $A(r)$ corresponding to the
conformally invariant case from Eq. (68) in $d=6$ and $s=\frac{3}{2}$, for a
set of $\eta $ parameters. Black hole formations with single / double
horizons are explicitly seen. Specifically, Fig. 3a for $0\leq \eta \leq
0.80 $ and Fig. 3b for $0.9\leq \eta \leq 1.0.$

\textbf{Table 1:} The table for $d$ versus $s$ that satisfies the condition $%
4s-d+2=0$. The reason for making this choice is technical for it simplifies
the expressions in (37) and (38) to great extend.

\bigskip

\bigskip

\end{document}